\documentclass[10pt]{jfm}
\usepackage{physics}
\usepackage[usenames, dvipsnames]{color}
\usepackage{ar}
\usepackage{amsmath,color,graphicx,amssymb}
\usepackage[font=small]{caption}
\usepackage{amsbsy}
\usepackage{latexsym}
\usepackage[normalem]{ulem}
\usepackage[english]{babel}
\usepackage{ifsym}
\usepackage{bbding}
\usepackage{wasysym}
\usepackage{latexsym}
\usepackage{epsfig} 
\usepackage{epsf,psfig}
\usepackage{epstopdf}
\usepackage{multirow}
\usepackage{array}
\usepackage{float}
\usepackage[english]{babel}
\usepackage{verbatim}
\usepackage{soul}
\usepackage[toc,page]{appendix}
\usepackage{setspace}
\usepackage{threeparttable}
\usepackage{booktabs,array}
\usepackage{color}
\usepackage{mathtools}
\usepackage{siunitx}
\usepackage{setspace}
\usepackage[numbers]{natbib}
\usepackage{authblk}
\usepackage{gensymb}
\usepackage{physics}
\usepackage[section]{placeins}
\definecolor{lightblue}{RGB}{48, 89, 155}
\definecolor{black}{RGB}{0, 0, 0}

\usepackage{float}
\def\bibfont{\footnotesize}
\usepackage{trimclip}

\usepackage{comment}

\begin{document}



\newtheorem{lemma}{Lemma}
\newtheorem{corollary}{Corollary}

\shorttitle{Lift reversal from vortex–surface phase coupling} 
\shortauthor{Qimin Feng et al.} 

\title{Lift reversal from vortex-surface phase coupling in a heaving foil near a free surface}

\author
 {
Qimin Feng\aff{1}
  \corresp{\email{qmfeng11@iastate.edu}},
Tianjun Han\aff{2},
Qiang Zhong\aff{1}
  }

\affiliation
{
\aff{1}
Department of Mechanical Engineering, Iowa State University, Ames, IA 50011, USA
\aff{2}
Department of Mechanical and Aerospace Engineering, University of Virginia, Charlottesville, VA 22904, USA
}

\graphicspath{{Figures/}}
\maketitle

\section{Abstract}
Classical descriptions of flapping propulsion near a free surface emphasize the energetic penalties of wave generation, treating the interface primarily as an energy sink. Here, we show that the same deformable boundary can also act as a phase-dependent kinematic constraint on vertical force generation. Using force measurements, particle image velocimetry and potential-flow simulations, we characterize how a free surface reorganizes vortex shedding for a heaving hydrofoil at moderate Reynolds number ($ O(10^4)$). For moderate to deep submergence, the cycle-averaged lift undergoes a systematic transition from repulsion to suction as the unsteady number increases. The reversal occurs within a narrow band of unsteady numbers, where the phase-shifted surface motion generates vertical advection that alters the pairing of trailing-edge vortices and redirects the wake momentum flux. A force decomposition shows that the reversal arises from a coordinated change in quasi-steady pressure loading and wake-induced force. These results identify the phase of the free-surface response, organized by unsteady number, as a key parameter governing near-surface lift and illustrate how deformable boundaries can reconfigure unsteady loading through vortex-surface phase coupling.

\section{Introduction}
Swimmers and bio-inspired robots operating near the free surface encounter a boundary that is compliant and time dependent, governed primarily by gravity-wave dynamics. When a body oscillates beneath it, the free surface not only absorbs energy through wave radiation but also responds with a frequency-dependent phase delay, particularly near the critical wave-excitation frequency \citep[e.g.][]{ellingsen2016waves}.

Previous studies have examined how free-surface excitation modifies thrust, power and efficiency in flapping foils. Surface deformation has been shown to reduce thrust, efficiency and modify the near-wake structure through wave–wake interaction \citep{zhu2006dynamics, cleaver2013periodically, chung2016propulsive, esmaeilifar2017hydrodynamic, deng2022effects}. Recent work has further shown that wake transport and wave–foil phase relations can influence propulsive performance under shallow submergence \citep{zheng2024free, ji2025hydrodynamics}. 

Despite these advances on streamwise performance and wave–wake interaction, the role of a deformable free surface in setting vertical loading remains comparatively underexplored. Rigid-wall ground-effect studies show that proximity to a solid boundary can reorganize vortex shedding and generate asymmetric vertical force through phase-sensitive interactions between the wake and the wall \citep{zhong2021aspect}. A free surface, however, is not a fixed boundary: it is a compliant interface whose motion has a frequency-dependent phase delay relative to the body. How such a moving boundary modifies wake trajectories, and whether this coupling can alter the sign of the cycle-averaged vertical lift, remains unknown. This question matters for near-surface swimmers (e.g. dolphins, penguins and surface-foraging fishes) that regulate vertical force to maintain depth, where even small imbalances can lead to unwanted surfacing or submergence \citep{blake2009biological}.

In this work, we characterize how a deformable free surface reorganizes vortex shedding to modify the cycle-averaged vertical force on a heaving foil. For moderate to deep submergence, we find the mean lift undergoes a systematic transition—switching from repulsion to suction—as the unsteady number $\tau$ increases. The transition occurs within a narrow band ($0.2 \lesssim \tau \lesssim 0.4$), where the phase-shifted surface motion produces vertical advection to alter the TEV pairing and redirect the wake momentum flux. A force decomposition shows that this change reflects a coordinated shift in the quasi-steady pressure field and the wake-induced impulse. Together, these results identify the phase of the free-surface response as a key parameter governing vertical loading near the interface.

\section{Methodology}

\subsection{Hydrodynamic force measurements}
\begin{figure}
    \centering
    \includegraphics[width=1\linewidth]{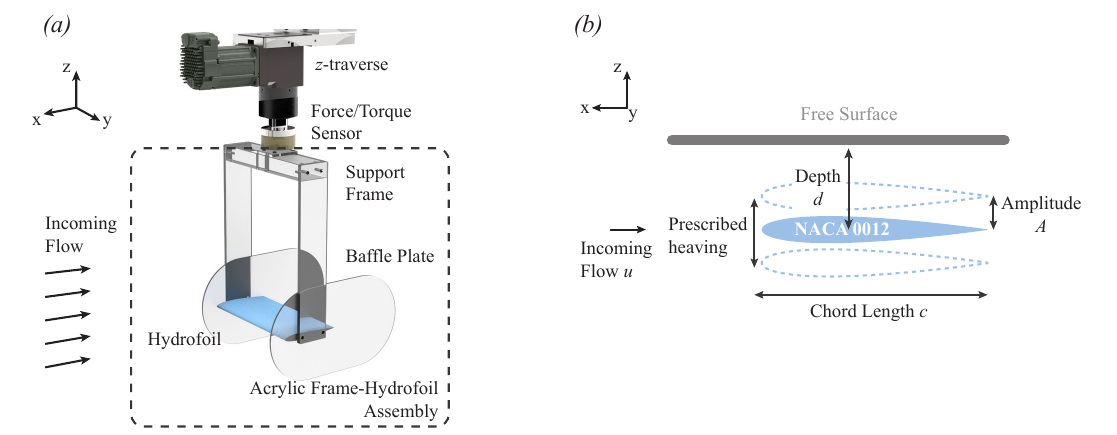}
    \vspace{-.2in}
    \caption{\textbf{Experimental setup.} (\emph{a}) Hydrofoil mounted along the z-axis traverse with end-plates to ensure two-dimensional flow. (\emph{b}) Schematic of the heaving kinematics ($U, A$) and geometric parameters ($d, c$) relative to the free surface.}
    \label{exp setup}
\end{figure}

Experiments were conducted in a closed-loop water tunnel (test section $3\times1\times0.75$ m) using a NACA-0012 hydrofoil (chord $c=95$ mm, span $2c$ with end-plates) driven in vertical heave motion $z(t)=A\sin(2\pi f t)$ (figure \ref{exp setup}).  Flow speed $U$ and free-surface proximity $d$ were varied to span a broad parameter space in unsteady number, Strouhal number and submergence ratio, characterized by 
$$
\tau=\frac{2\pi fU}{g},\qquad St=\frac{2fA}{U},\qquad d/c,
$$
where $g$ is gravitational acceleration. We surveyed $St=0.2$--$0.8$, $\tau=0.1$--$1.0$, and $0.31\le d/c\le2.3$ at fixed $A/c=0.15$ ($Re \approx 0.7$--$4.4\times 10^4$), with each of the 840 cases repeated five times. Lift forces ($L$) were sampled at 200 Hz with a six-axis load cell (ATI Mini40 IP65: SI-80-4), normalized as $C_L = L/(\rho c^2 U^2)$ ($\rho$ is water density), and cycle-averaged over 20 periods. Static tares (frame weight and buoyancy) were measured using an ultra-low range load cell (Interface) and subtracted to yield net hydrodynamic forces.

\subsection{PIV measurements of wake fields and surface deformation}
Two-dimensional Particle Image Velocimetry (PIV) captured wake structures using $20\,\mu$m polyamide seeding particles illuminated by a 10\,W laser sheet (Lumivanta-LaserWave). Images ($4096 \times 2304$ px) were recorded by a high-speed camera (Photron Nova R3-4K) synchronized with the hydrofoil motion. We phase-averaged 600 frames over 10 cycles to obtain 60 representative phases per period. Vector fields were computed using a cross-correlation algorithm (LaVision Davis 10) with $64 \times 64$ px overlapping interrogation windows. Surface profiles were simultaneously extracted from the high-speed recordings to define the boundary conditions for simulations and wake visualizations. Local wave-induced velocities were estimated from the temporal evolution of the tracked surface points using the kinematic free-surface condition.

\subsection{Force decomposition via potential-flow simulation}
To diagnose the forces contributions behind the foil oscillating near the free surface, we conducted two-dimensional potential-flow simulations by matching the exact experimental conditions. The free surface was explicitly modeled as a moving vortex sheet, and the free-surface boundary condition is prescribed using interpolated experimental wave data. This approach intentionally decouples the force decomposition from fully coupled wave generation, allowing the influence of a given free-surface kinematics on the hydrodynamic force to be isolated. At each time step in the simulation, no-flux boundary condition is enforced on the foil and the moving surface to solve for the flow field and unsteady forces. Following the classic lift decomposition \citep{von1938airfoil}, the instantaneous lift from the simulations is decomposed into the added-mass ($C_L^{\mathrm{AM}}$), quasi-steady ($C_L^{\mathrm{QS}}$), and wake-induced ($C_L^{\mathrm{WI}}$) components. See \citep{han2024revealing} and \citep{zhu2025wavenumber} for details on the numerical and force decomposition methods. Although the added-mass force $C_L^{\mathrm{AM}}$ oscillates in time, its cycle-averaged contribution effectively vanishes for the periodic motions considered here \citep{han2024revealing,zhu2025wavenumber}.  Consequently, the mean lift depends solely on the remaining components: $\overline{C}_L = \overline{C}_L^{\mathrm{QS}} + \overline{C}_L^{\mathrm{WI}}$ and the added-mass term does not contribute to the polarity reversal.

\section{Results}
\begin{figure}
    \centering
    \includegraphics[width=1\textwidth]{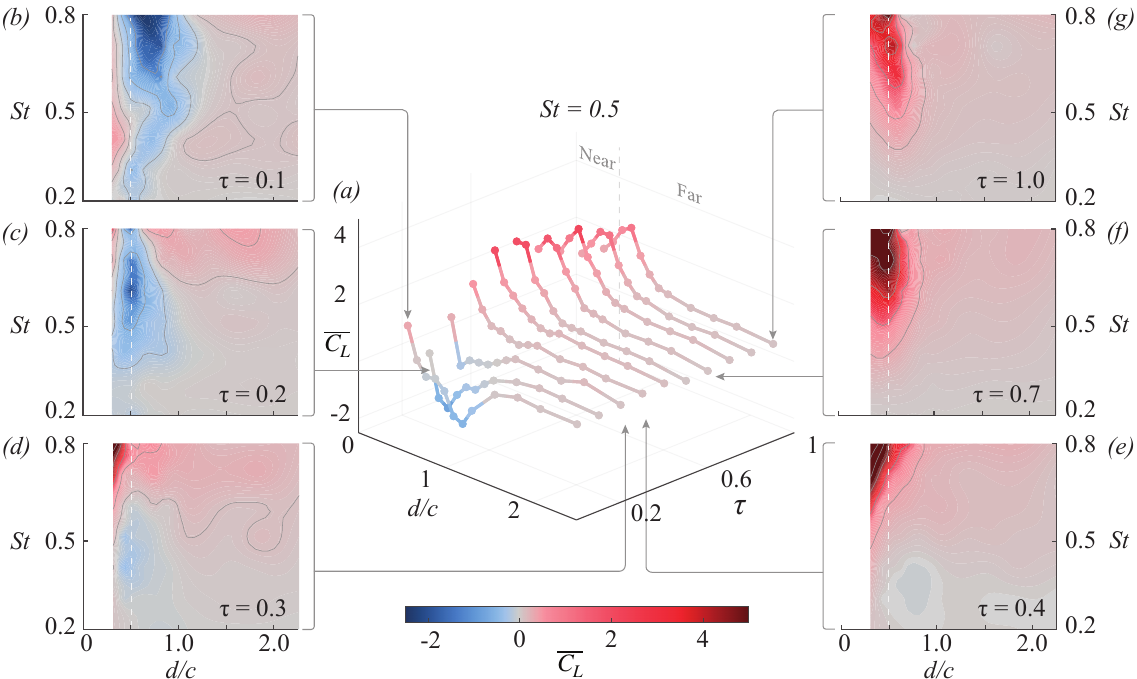}
    \caption{\textbf{Time-averaged lift coefficient ($\overline{C}_L$) regimes defined by the unsteady number $\tau$, Strouhal number $St$, and submergence $d/c$.}
    (\emph{a}) Lift polarity reversal at $St=0.5$, showing the transition from repulsion (negative) to suction (positive) with increasing $\tau$. 
    (\emph{b}--\emph{g}) Contours of $\overline{C}_L$ in the $(St, d/c)$ domain for fixed $\tau$. The dashed vertical lines in (\emph{a}--\emph{g}) delineate the near-surface and far-field regions. Blue is repulsion and red is suction.
    }
\label{fig 2 : asymmetric lift}
\vspace{-15pt}
\end{figure}

\subsection{Mean lift reversal across unsteady number}
In general, a heaving hydrofoil operating near a free surface generates a nonzero mean lift $\overline{C}_L$, which exhibits a clear lift reversal governed primarily by the unsteady number $\tau$, with secondary modulation from the Strouhal number and submergence.

Figure~\ref{fig 2 : asymmetric lift}(\textit{a}) maps the landscape of $\overline{C}_L$ at a representative Strouhal number ($St = 0.5$), revealing that $\tau$ organizes the overall polarity of the vertical force. At low unsteady numbers ($\tau \le 0.2$), the mean lift is predominantly negative (depth-directed), indicating a repulsive force pushing the foil away from the free surface. Conversely, at higher unsteady numbers ($\tau \ge 0.4$), the polarity reverses: the lift becomes consistently positive (surface-directed), indicating a suction force pulling the foil toward the free-surface. For the intermediate regime ($0.2 \lesssim \tau \lesssim 0.4$), the force fluctuates around zero, marking a transition between the repulsive and suction regimes. Beyond polarity, submergence $d/c$ modulates the interaction strength: the lift magnitude $|\overline{C}_L|$ generally peaks at intermediate depth ($d/c \approx 0.5$) before decaying as surface effects weaken.

The fixed-$\tau$ contours (figure~\ref{fig 2 : asymmetric lift}(\textit{b}–\textit{g})) further illustrate the modulating role of Strouhal number. At $\tau = 0.1$ (figure~\ref{fig 2 : asymmetric lift}\textit{b}), the lift is negative across nearly the entire $(St, d/c)$ plane, with the strongest repulsion occurring near $d/c \approx 0.75$. The peak magnitude scales with $St$, while the negative-lift region broadens. As submergence departs from this band---either toward very shallow immersion or toward deeper submergence---the mean force decays toward zero.

This spatial pattern at $\tau = 0.1$ shares a superficial visual similarity with unsteady ground effect: a strong negative-lift pocket that strengthens with $St$ and decays away from the boundary \citep{quinn2014unsteady,zhong2021aspect}. A comparable near-surface enhancement has also been reported for low–Re swimmers near a free surface, where proximity produced a measurable thrust boost \citep[$Re \sim O(10^2)$;][]{zheng2024free}. In the present $Re \sim O(10^4)$ regime, however, the resemblance is limited. Our measurements indicate only a weak thrust increase at low $St$ ($\lesssim 0.3$), and the lift curves presented here show no indication of the plateau or equilibrium-like features characteristic of rigid-wall ground effect.

The transition toward the high-$\tau$ regime occurs within the narrow transition band ($0.2 \lesssim \tau \lesssim 0.4$), where the free-surface response varies rapidly with Strouhal number. In this band (figure~\ref{fig 2 : asymmetric lift}(\textit{d,e})), the mean lift curves change sign rapidly: at $\tau = 0.3$, the force retains a downward bias at lower Strouhal numbers ($St \lesssim 0.6$), but increasing $St$ drives a reversal to positive lift; by $\tau \approx 0.4$, positive lift dominates most of the $(St, d/c)$ plane and the transition threshold shifts to lower $St$ ($St \lesssim 0.35$).

For $\tau > 0.4$, the transition is largely complete, with positive lift dominating the $(St, d/c)$ domain (figure~\ref{fig 2 : asymmetric lift}\textit{f,g}). The region of suction expands with increasing unsteady number and consolidates around $d/c \approx 0.5$. Notably, at the smallest submergence values ($d/c = 0.31$), the lift attenuates relative to the peak, creating a distinct band of reduced force very close to the free surface. This high-$\tau$ topology demonstrates the establishment of the upward-bias regime identified in figure~\ref{fig 2 : asymmetric lift}(\textit{a}), and illustrates how the strength and spatial extent of the suction region increase as $\tau$ moves beyond the transition band.

\begin{figure}
    \centering
    \includegraphics[width=0.95\textwidth]{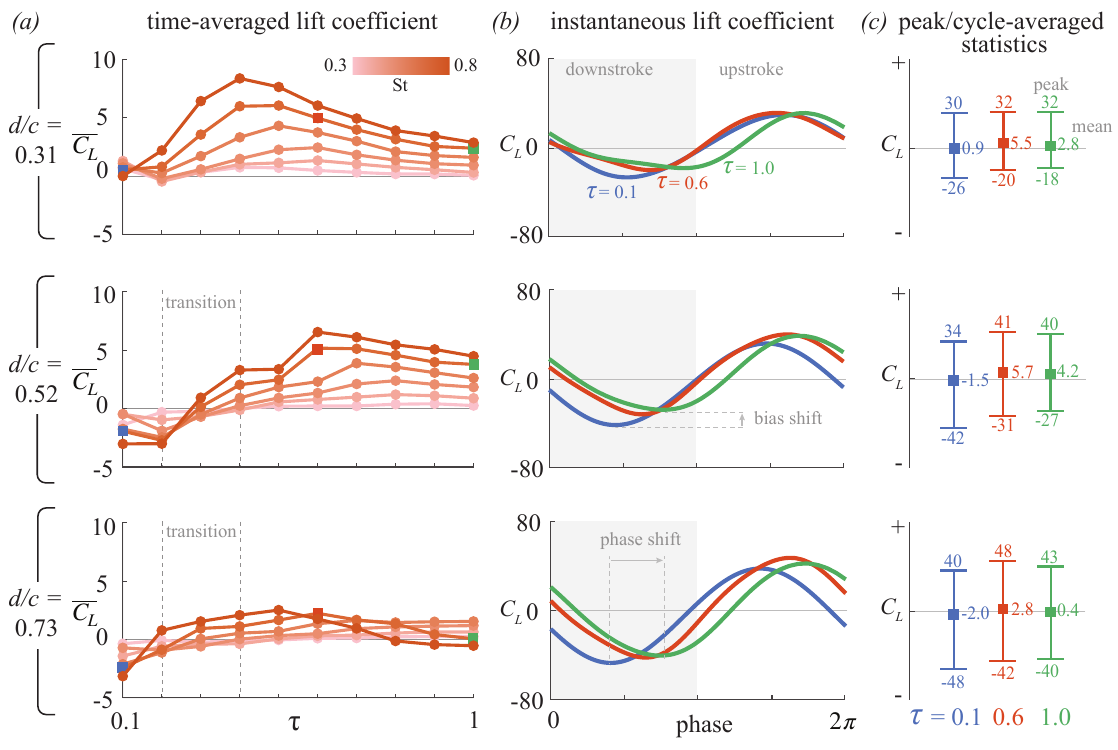}
    \vspace{-5pt}
    \caption{\textbf{Cycle-resolved lift ($C_L(t)$) dynamics governing the mean-lift reversal.} (\textit{a}) $\overline{C}_L$ versus $\tau$ across varying $St$ and submergence depths. Dashed vertical lines mark the transition band; square markers indicate the specific cases analyzed in (\textit{b}) and (\textit{c}). (\textit{b}) Instantaneous lift waveforms $C_L(t)$ for representative cases ($St=0.7$), highlighting the distinct bias shift and phase shift mechanisms. (\textit{c}) Peak-to-peak statistics showing the maximum, minimum, and cycle-averaged (square marker) lift.}
\label{fig 3 CL_tau}
\vspace{-10pt}
\end{figure}
\subsection{Instantaneous waveforms underlying the $\overline{C}_L$ reversal}
Figure~\ref{fig 3 CL_tau}(a) plots the mean lift $\overline{C}_L$ against $\tau$ for three
submergence distances, providing a resolved view of the interplay between $\tau$, $St$, and
$d/c$ within the transition band. 

For moderate to deep submergence ($d/c \ge 0.52$), the transition identified in §4.1 occurs within a narrow interval ($0.2 \lesssim \tau \lesssim 0.4$): the mean lift switches from negative (repulsion) to positive (suction).  Within this band, increasing $St$ is associated with a larger force magnitude and a shift of the transition threshold to lower $\tau$; for example, strong heaving ($St = 0.8$) enters the positive-lift regime earlier than weaker heaving ($St = 0.4$). This trend indicates that the transition reflects not only the timescale competition captured by $\tau$, but also variations in the amplitude of vortex shedding with Strouhal number.

At the smallest distance tested ($d/c = 0.31$), the transition becomes substantially muted: nearly all curves exhibit a positive mean lift, and the dependence on $\tau$ weakens, indicating that this extreme near-surface case no longer follows the $\tau$-ordered lift reversal observed at larger submergence.

The corresponding instantaneous waveforms $C_L(t)$ (figure~\ref{fig 3 CL_tau}(\textit{b})) reveal kinematic changes associated with these trends. Here, we focus on the high-$St$ regime where the reversal mechanism is most pronounced. 
For $d/c \ge 0.52$, the waveform retains approximate symmetry about the mean; the shapes of the downstroke and upstroke segments remain largely invariant with $\tau$. Instead, the polarity reversal in $\overline{C}_L$ arises primarily from a systematic bias shift (see figure~\ref{fig 3 CL_tau}(\textit{b})): the entire waveform translates vertically from a negative to a positive baseline as the unsteady number increases. Concurrently, a phase shift is observed in the waveform peaks. This dual variation is consistent with a systematic reorganization of the unsteady loading as the phase relation between the free-surface and the foil motion varies with increasing $\tau $.

At $d/c = 0.31$, however, the waveform exhibits strong asymmetry that persists across $\tau$. The downstroke becomes increasingly flattened at higher $\tau$, whereas the upstroke peak remains sharp. These distortions differ qualitatively from the phase-driven changes at larger depth and highlight the dominance of near-field confinement when the foil operates very close to the interface.

Figure~\ref{fig 3 CL_tau}(\textit{c}) quantifies waveform variations through peak-to-peak and mean force metrics. As the foil approaches the free surface ($d/c \to 0.31$), the cycle-averaged lift magnitude $|\overline{C}_L|$ (square markers) increases slightly, yet the peak-to-peak amplitude (total vertical range) decreases.

\subsection{Wake dynamics near the free surface}

\begin{figure}
    \centering
    \includegraphics[width=0.95\textwidth]{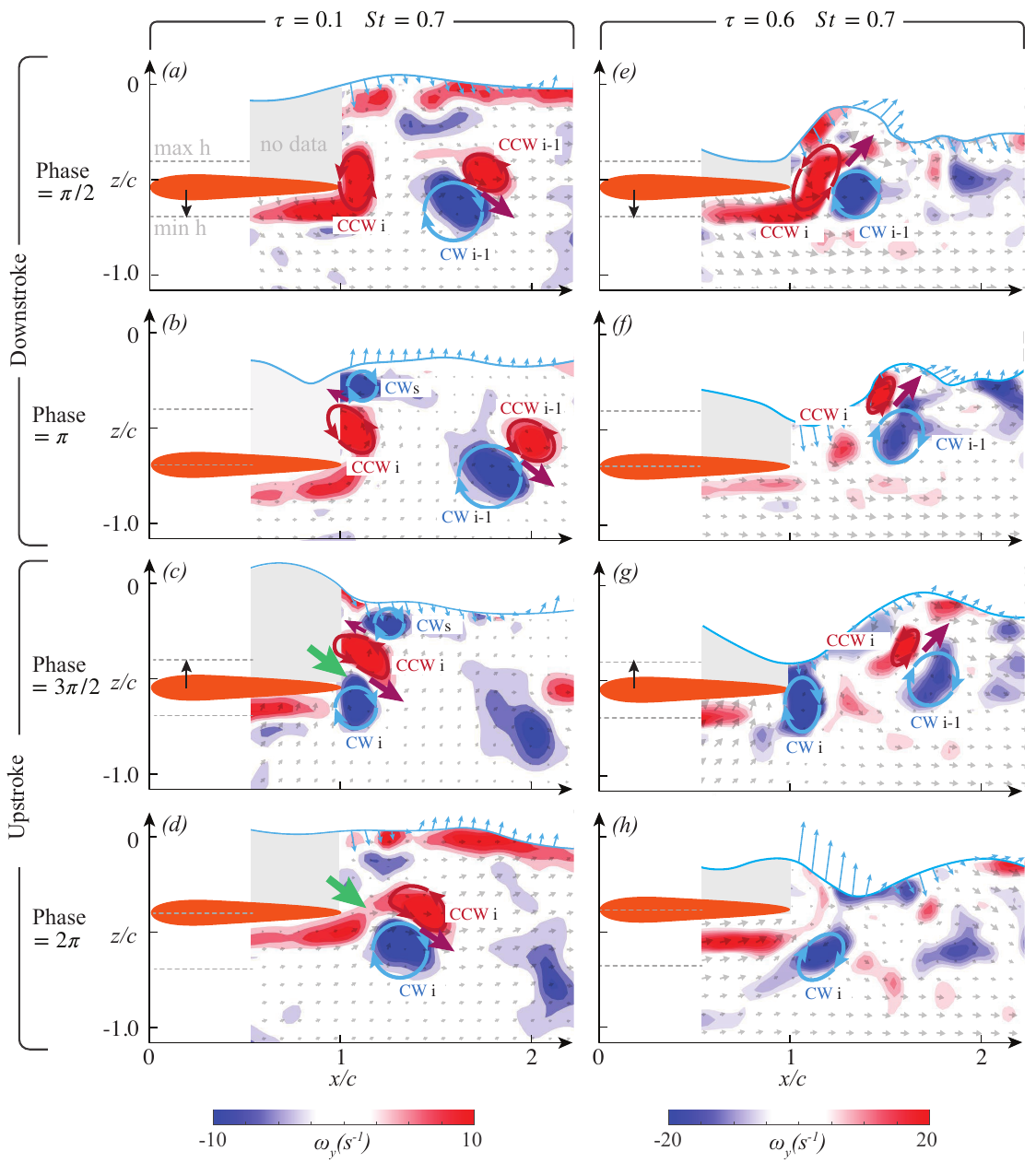}
    \vspace{-5pt}
    \caption{\textbf{The unsteady number $\tau$ modulates the interaction between the free surface and trailing-edge vortices (TEVs), leading to distinct deflected wakes.} Panels (\emph{a--d}) correspond to $\tau = 0.1$ and $St = 0.7$; panels (\emph{e--h}) correspond to $\tau = 0.6$ and $St = 0.7$, both at $d/c = 0.52$. Spanwise vorticity $\omega_y$ is shown in colour (red: CCW; blue: CW). Annotated vectors isolate distinct transport mechanisms: Light-blue arrows denote surface-wave--induced velocity; green arrows indicate pressure-gradient forcing from the foil--surface geometry; and purple arrows represent the self-induced motion of the TEV pair.}
    
    \label{fig 4 wake}
    \vspace{-16pt}
\end{figure}

To exam wake dynamics underlying the lift reversal, we analyze the phase-resolved vorticity fields for representative low- and high-$\tau$ cases (figure~\ref{fig 4 wake}). For a low unsteady number ($\tau = 0.1$; figure~\ref{fig 4 wake}(\textit{a--d})), the free surface exhibits only small, rapidly adjusting deformations in this regime. As the hydrofoil reaches its maximum downward velocity (phase $\pi/2$), the velocity field associated with the local surface deformation exhibits a downward component above the trailing edge. This produces a confinement that limits the vertical development of the newly shed CCW$_i$ vortex, keeping it close to the trailing edge (figure~\ref{fig 4 wake}(a)). As the motion continues, the expanding CCW$_i$ interacts with the interface and generates a weak clockwise secondary vortex (CW$_s$) due to surface curvature and kinematic constraints \citep{ohring1991interaction,sarpkaya1996vorticity}; the induced velocity of this CCW$_i$--CW$_s$ pair further retards the downstream drift of CCW$_i$ (figure~\ref{fig 4 wake}(\textit{b})).

When the foil reverses and reaches its maximum upward velocity (phase $3\pi/2$; figure~\ref{fig 4 wake}(\textit{c})), the free-surface crest above the mid-chord has begun to descend while the foil is still rising. This relative motion compresses the fluid between the foil and the free-surface, consistent with a locally elevated pressure above the mid-chord of the foil \citep{deng2022effects}. The newly shed clockwise vortex (CW$_i$) then encounters the delayed CCW$_i$, forming a dipole whose self-induced velocity (purple arrows) directs the pair downward and slightly downstream. This dipole motion aligns with—and reinforces—the pressure-gradient forcing (green arrows), yielding a downward-oriented jet (figure~\ref{fig 4 wake}(\textit{d})).

For high unsteady number ($\tau = 0.6$; figure~\ref{fig 4 wake}(e--h)), the surface
deformation exhibits a pronounced temporal offset relative to the foil motion, placing the surface crest near the vortex-shedding phase. As the foil reaches maximum downward velocity (phase $\pi/2$; figure~\ref{fig 4 wake}(\textit{e})), the delayed surface crest induces a strong upward--rightward velocity (light-blue arrows). This flow immediately lifts the newly shed counter-clockwise vortex (CCW$_i$) toward the interface, positioning it significantly closer to the free surface than in the low-$\tau$ case. As a result, CCW$_i$ pairs with the preceding clockwise vortex (CW$_{i-1}$) at a noticeably earlier phase. The resulting dipole, reinforced by the velocity field associated with the surface crest, produces a persistent upward-directed jet (figure~\ref{fig 4 wake}(\textit{e--h})).

As the foil reverses and approaches its maximum upward velocity (phase $3\pi/2$; figure~\ref{fig 4 wake}(\textit{g})), the newly shed clockwise vortex (CW$_i$) remains trapped near the trailing-edge region. This occurs because, unlike in deeper-water cases, the strong surface-induced advection rapidly removes its counter-rotating partner upward (the CCW$_i$), leaving CW$_i$ without an opposing vortex to promote downstream convection. This delayed CW$_i$ subsequently pairs with the next counter-clockwise vortex (CCW$_{i+1}$) in the following cycle, forming the next surface-directed dipole.

Taken together, the wake visualizations indicate that the lift reversal is associated with a reorganization of the wake topology: from surface-constrained downward pairing at low $\tau$ to wave-advected upward deflection at high $\tau$. While these patterns qualitatively rationalize the lift reversal, vorticity fields alone cannot isolate the relative contributions of quasi-steady and wake-induced loads. To quantify these dynamical origins, we perform a potential-flow force decomposition in §4.4.

\label{section: PIV wake analyze}

\subsection{Simulation and force decomposition}

\begin{figure}
    \centering
    \includegraphics[width=0.95\textwidth]{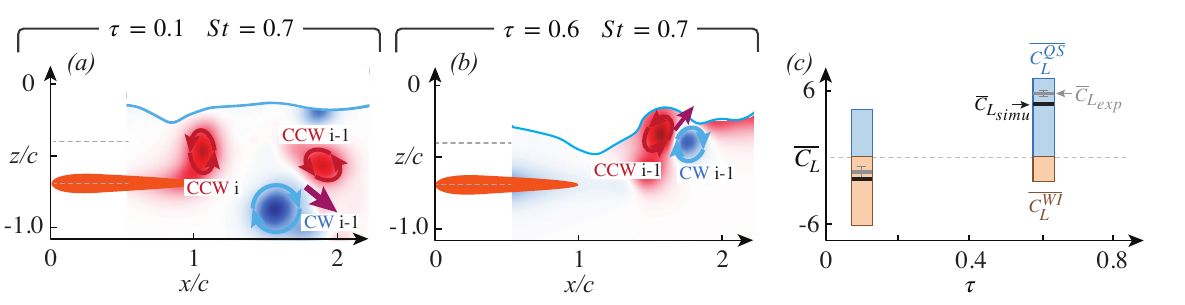}
    \vspace{-5pt}
    \caption{\textbf{Simulation wake fields and cycle-averaged lift decomposition.} (\emph{a,b}) Simulated wake topology at phase $\pi$ reproducing the shift from a downward dipole (low $\tau$) to an upward jet (high $\tau$). (\emph{c}) Decomposition of mean lift. The theoretical reconstruction (black lines) captures the experimental lift reversal (grey lines), driven by the concerted shift of quasi-steady ($\overline{C}_L^{QS}$) and wake-induced ($\overline{C}_L^{WI}$) components.}
    \label{fig 5 simu}
    \vspace{-10pt}
\end{figure}
To isolate the dynamical origins of the lift reversal, we employ the potential-flow solver as a diagnostic tool. By prescribing the experimentally measured surface deformation as a kinematic boundary condition, we intentionally decouple the force decomposition from the complexities of fully coupled wave generation. While this approach does not predict the surface evolution, it ensures kinematic consistency with the physical reality. Consequently, the fact that the simulations reproduce the experimental wake topology and forces, specifically the shift from a downward (low $\tau$) to an upward-deflected jet (high $\tau$), demonstrates that the potential-flow model correctly captures the kinematic rearrangements of the TEV that correlate with the sign of the wake-induced lift, despite the absence of viscous diffusion and minor secondary structures (e.g., the CW$_s$ vortex observed in figure~\ref{fig 4 wake}(\textit{b})). The quantitative agreement between this 2D model and the finite-span experiments suggests that three-dimensional effects---including tip vortices and transverse wave radiation---are secondary to the planar wake reorganization mechanism described here. Since the cycle-averaged added-mass force is negligible for these periodic kinematics, the mean lift is governed entirely by the balance between quasi-steady ($\overline{C}_L^{\mathrm{QS}}$) and wake-induced ($\overline{C}_L^{\mathrm{WI}}$) contributions.

Figure~\ref{fig 5 simu}(\textit{c}) reveals that the mean lift reversal arises from a collaborative mechanism involving both terms. First, $\overline{C}_L^{\mathrm{QS}}$ increases with $\tau$, consistent with a growing cycle-wise asymmetry in the quasi-steady pressure field and with free-surface-induced pressure biases reported for a plunging foil \citep{esmaeilifar2017hydrodynamic}. As the heaving motion increasingly outpaces the adjustment of the free surface at larger $\tau$, the interface no longer follows the foil quasi-statically but develops a bulged elevation directly above it (figures~\ref{fig 4 wake} and \ref{fig 5 simu}(\textit{a,b})). This deformation leads to compression of the fluid between the foil and the surface and creates a persistent high-pressure lobe above the foil, biasing the quasi-steady pressure difference between the upper and lower surfaces and thereby increasing $\overline{C}_L^{\mathrm{QS}}$. This effect produces a noticeable upward bias and phase shift in the cycle-averaged lift. Second, $\overline{C}_L^{\mathrm{WI}}$ undergoes substantial mitigation. While the downward-oriented dipole at low $\tau$ generates a large negative wake-induced lift, this negative force is reduced substantially at high $\tau$, consistent with the rearranged TEV pairing and upward deflection.

Crucially, the decomposition reveals that the lift reversal is not a singular event but a composite transition. The reversal requires the concerted shift of two mechanisms: the quasi-steady pressure bias (geometric phase) provides the necessary positive lift, while the wake reorganization (topological phase) removes the strong downward suction. This indicates that $\tau$ governs the lift reversal by simultaneously regulating both the boundary pressure field and the vortex trajectory.

\section{Discussion and Conclusion}
The present study highlights a form of phase-sensitive interaction in flapping-foil hydrodynamics. We demonstrate that the relative timing between surface-wave adjustment and vortex shedding reorganizes the dipole pairing sequence, producing a lift reversal that is well characterized by the unsteady number $\tau$ across the explored parameter range. Whereas prior studies emphasized the energetic consequences of free-surface excitation— notably the additional drag associated with wave radiation \citep{zhu2006dynamics,chung2016propulsive,esmaeilifar2017hydrodynamic}---the present results highlight a complementary aspect of the same interaction: the timing of surface deformation acts as a dynamic constraint that shapes vortex transport. The force decomposition shows that the polarity reversal arises from a coordinated shift in two components: a developing quasi-steady pressure asymmetry and a reorganization of the wake-induced force as the TEV pairing sequence changes.       

This $\tau$-controlled lift reversal is expressed most clearly at intermediate submergence depths ($d/c \gtrsim 0.5$), where the mean-lift curves collapse when plotted against $\tau$. At the smallest gap tested ($d/c = 0.31$), however, this collapse breaks down: the lift varies only weakly with $\tau$ and no reversal occurs. These observations suggest that geometric confinement dominates the interaction very near the surface, suppressing the phase-governed reorganization of TEV trajectories that drives the lift reversal at larger depths.

The transition band $0.2 \lesssim \tau \lesssim 0.4$ overlaps with the range previously associated with noticeable changes in thrust production for plunging foils in the low-$St$ regime \citep{zhu2006dynamics,cleaver2013periodically}. In those low-$St$ cases, wave growth and associated drag peaks dominate the overall performance, whereas the higher-$St$ regime explored here displays its primary signature through phase-dependent reorganization of the TEV topology. Within this window, the free surface exhibits the strongest sensitivity to the imposed frequency, which modifies the vertical position and relative timing of the shed vortices. This interpretation is consistent with recent observations that wake topology in flapping foils is strongly conditioned by phase relations rather than solely the amplitude of surface deformation \citep{samsam2024fluid,ji2025hydrodynamics}. Accordingly, the lift reversal documented here can be viewed as a particular manifestation of this broader phase-synchronization behavior.

The central role of $\tau$ emphasizes that the lift transition is fundamentally tied to a competition of timescales: the surface-adjustment time and the TEV-shedding period \citep{grue1988propulsion}. Neither Froude number ($Fr$), which characterizes advective balance, nor Strouhal number ($St$), which sets circulation strength \citep{triantafyllou1993optimal}, alone captures the temporal structure of this interaction. Instead, $\tau$ isolates the condition under which the phase of surface motion modifies the kinematics of vortex release and pairing: as $\tau$ is varied, the relative phase between free-surface elevation and foil motion shifts, so that surface-induced vertical velocities either advance or delay TEV transport and pairing (see figure~\ref{fig 4 wake}). Within this framework, $St$ and $d/c$ act as modulators—affecting the strength of the coupling and the precise transition threshold—but the overall bifurcation-like trend follows the ordering set by $\tau$. A complete predictive scaling for the magnitude of TEV–surface coupling likely requires a more systematic sweep of $(\tau, St, d/c)$ and a dedicated theoretical model, which lies beyond the scope of the present work.

This phase-organized perspective broadens the classical view of how deformable interfaces mediate unsteady propulsion. It shows that the free surface can reconfigure unsteady lift by altering the phase relation between surface deformation and shedding, without requiring changes in stroke asymmetry. This suggests that biological and robotic swimmers may leverage surface proximity to modulate vertical force through frequency-based timing control. More broadly, the emergence of a $\tau$-organized wake topology highlights that deformable boundaries introduce additional dynamical clocks into vortex-dominated flows, enabling force regimes not observed near rigid walls.

\bibliographystyle{jfm}
\bibliography{references}
\vspace{10pt}
\textbf{Acknowledgments} This work is funded by Link Foundation and Iowa State University.
\vspace{50pt}
\textbf{Declaration of interests}
The authors report no conflict of interest.

\FloatBarrier

\end{document}